\documentclass[11pt,twoside]{article}
\usepackage{./asp2014}

\aspSuppressVolSlug
\resetcounters

\begin{document}

\title{{\sc XTgrid \textup{Live:}} Online spectral analyses with {\sc Tlusty} models}
\author{P\'eter N\'emeth$^{1,2}$}
\affil{$^1$Astronomical Institute of the Czech Republic, 25165 Ond\v{r}ejov, Czech Republic}
\affil{$^2$Astroserver.org, 8533 Malomsok, Hungary\\
\email{peter.nemeth@astroserver.org}}

\paperauthor{Peter Nemeth} {peter.nemeth@astroserver.org} {0000-0003-0963-0239}{Astronomical Institute of the Czech Republic} {Stellar Department} {Ondrejov} {} {25165} {Czech Republic}

\begin{abstract}
Non-LTE model atmospheres are in the forefront of quantitative spectral analyses of early type stars. Yet, such models are barely used in large surveys or even for individual systems. The major reason for this is likely the extreme calculation requirements to construct such physically realistic models. To relieve this, we have developed a web-service for the spectral analysis program {\sc XTgrid}, which is meant to be a handy interface to {\sc Tlusty}. Having the procedure available from a web browser makes spectral analyses a routine.
\end{abstract}

\section{A brief introduction to {\sc XTgrid}}
The development of our spectral analysis pipeline for {\sc Tlusty} (\citealt{hubeny95, hubeny17}) has started a decade ago. Its first science results were published from a homogeneous modeling of a large set of hot subdwarf optical spectra \citep{nemeth12}. This task required the program to oversee the whole analysis procedure, including the ability to recover from convergence failures, adjust the model details, change model atoms as ionization equilibrium requires, etc. Scalability for parallel computing was important from the beginning, as well as a high degree of generality to solve any observation the program may get and {\sc Tlusty} can reproduce. For such reasons {\sc XTgrid} was designed to calculate everything on-the-fly, models on demand, and be applicable for a range of very diverse spectra regarding coverage, resolution or quality. 

The fitting procedure begins with a quick estimate of the spectral type and the iterative steepest-descent chi-square minimization procedure continues from this initial model. By applying successive approximations along the steepest gradient of the global chi-square, the procedure converges on the best fit. 

Because {\sc XTgrid} does not require a pre-calculated grid of models it can address even the most peculiar stars using off-the-shelf desktop computers. To accelerate fitting, a model cloud is built during successive runs and previously calculated models are recycled in future runs. The parameter space coverage of this cloud is quite irregular following the preference of nature and due to the large number of parameter combinations. 
These features make {\sc XTgrid} flexible and suitable for a web-service. Next, we designed an online interface, {\sc XTgrid} Live, which allows to start a spectral analysis as easy as to upload an observation and click a button. 
Although such a black-box like use of model atmosphere codes is discouraged, it helps reaching a larger audience and far more applications where the non-LTE approach is necessary. 
As the source of {\sc Tlusty} is well documented and publicly available, interested users can find all details online\footnote{\url{http://nova.astro.umd.edu/}}. 

\section{Target audience}
A spectral analysis web-service is quite useful to observers, who need a real-time evaluation of incoming spectra, as well as to researchers working with different codes and want to make comparisons. Such a service is also useful to students who need a quick start in a new project.  

\section{Applications}
\subsection{Spectroscopy of large samples}
Large spectroscopic surveys, such as the LAMOST survey, are excellent resources to search for homogeneous observational data for any stellar types. Such data sets are best analyzed with global methods to exploit their homogeneity and reduce systematics in the results. 
Fig. \ref{fig1} shows a sample of \textasciitilde 1000 stars from LAMOST DR5 \citep{lei18}. {\sc Tlusty/XTgrid} was used to spectroscopicly confirm hot subdwarf candidates selected from Gaia DR2 based on astrometric and photometric data.

\subsection{Binary spectral decomposition}

\begin{figure}
\begin{minipage}{0.49\textwidth}
  \centering
  \vspace{-4pt}
  \includegraphics[width=\textwidth]{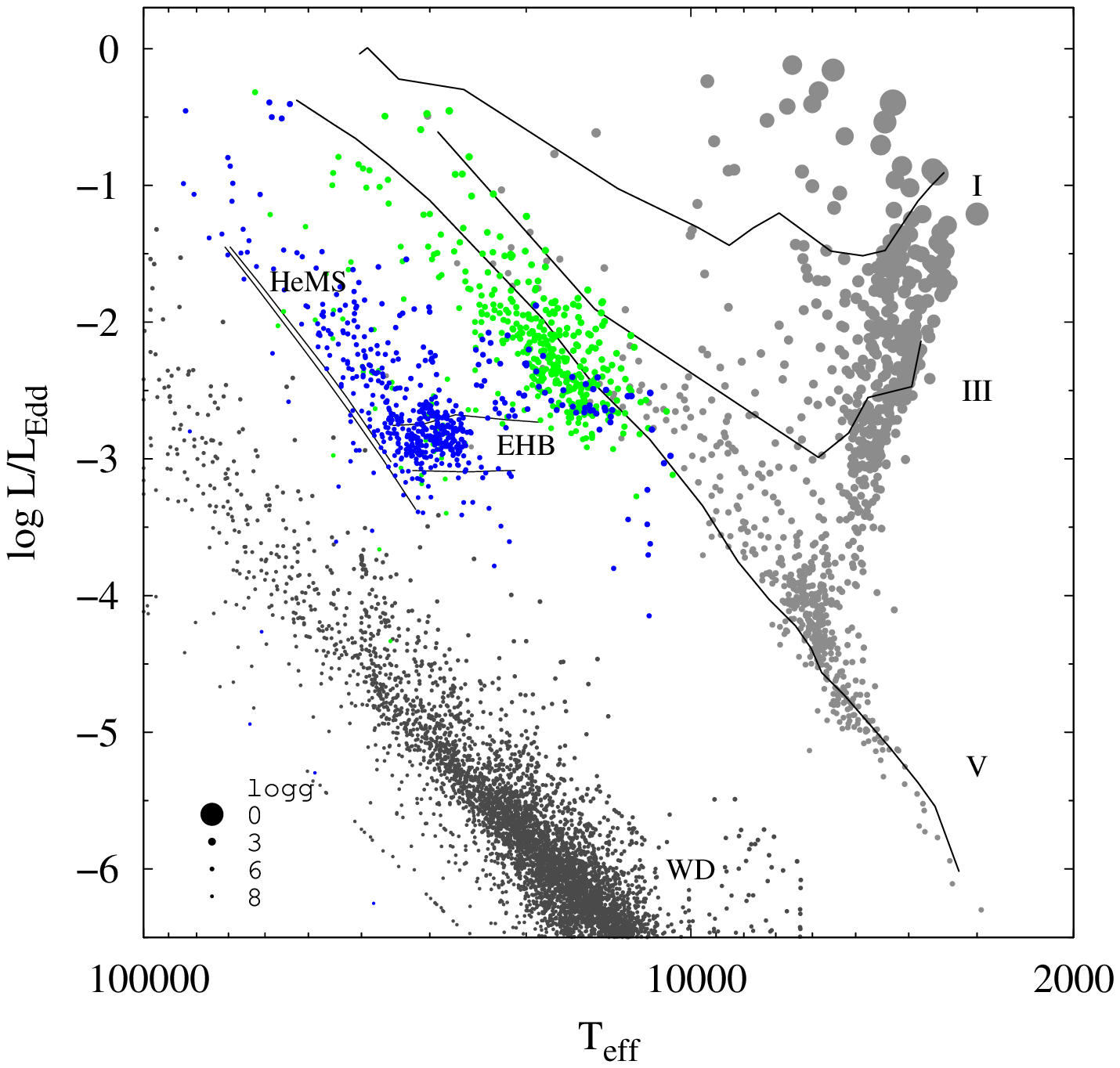}
  \caption{Hot subdwarf (blue) and upper main sequence star candidates (green) from LAMOST DR5, based on Gaia DR2 data.\label{fig1}}
\end{minipage}%
\centering
\begin{minipage}{0.5\textwidth}
  \vspace{-4pt}
  \centering
  \includegraphics[width=\textwidth]{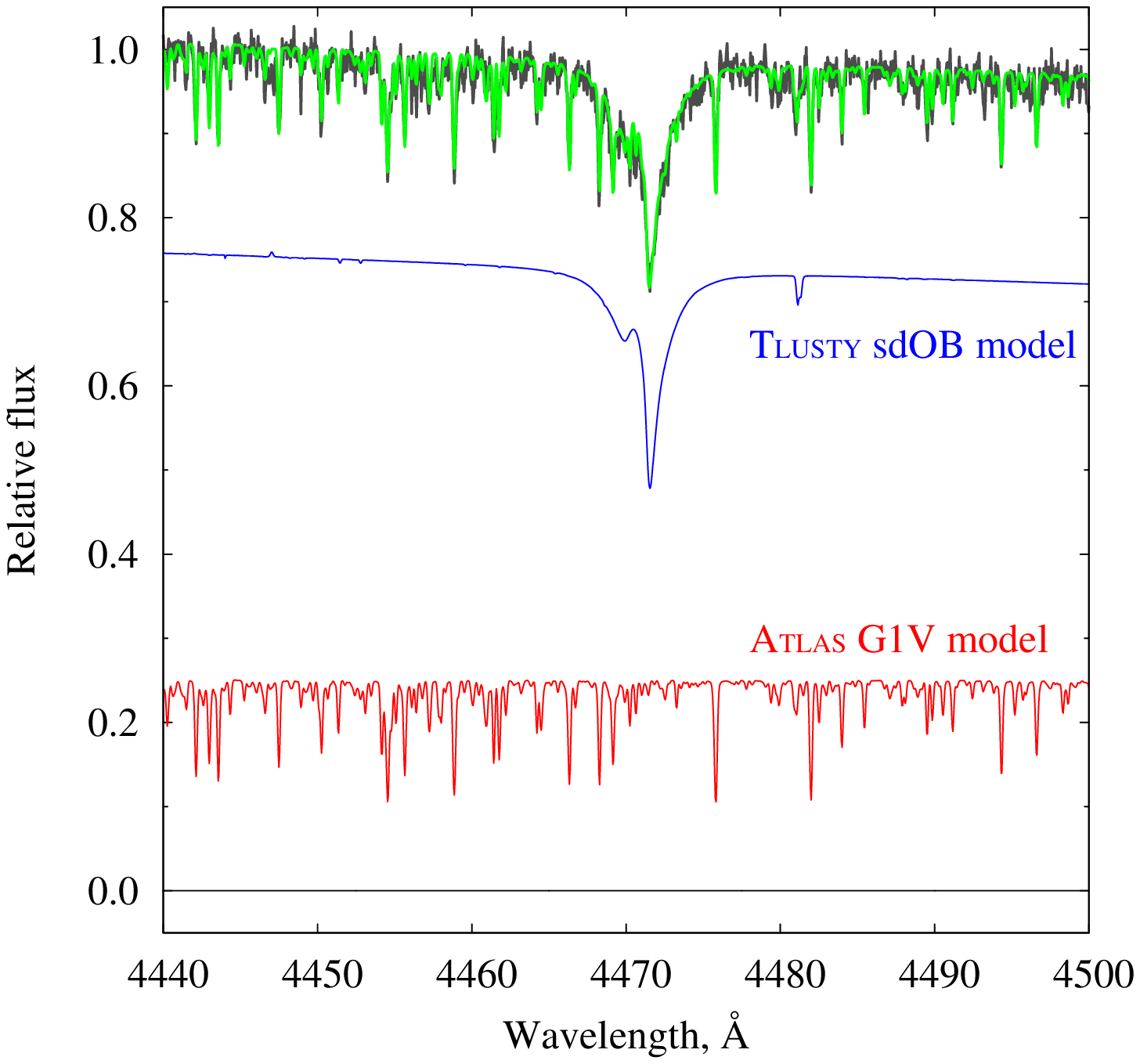}
  \caption{
  Spectral decomposition of the sdOB+G1V binary SB\,744. 
  The sum of the components (green) fits the observation (black). 
  \label{fig2}}
\end{minipage}
\end{figure}

An iterative spectral decomposition algorithm is included in {\sc XTgrid} that allows a characterization of composite spectra double lined binaries (SB2). 
{\sc XTgrid} applies LTE spectra from public libraries for the cool companion and {\sc Tlusty/Synspec} non-LTE spectra for the hot component of the binary. 
The method is based on an iterative reconstruction of the composite spectrum from a linear combination of two stellar models. 
This allows a decomposition from a single spectrum without an extensive coverage of the radial velocity curve. Fig. \ref{fig2} shows the spectral decomposition of the sdOB+G1V type binary SB\,744 from \cite{vos18}. 

The high level of degeneracy among the stellar and binary parameters is a major problem in decomposition. To overcome these, we use a combination of flux calibrated low-resolution spectra and high resolution data to infer the individual stellar properties as well as their contributions to the composite spectrum.

\subsection{Simultaneous Ultraviolet-Optical-Infrared (UVOIR) spectral modeling}
\articlefigure[width=\textwidth]{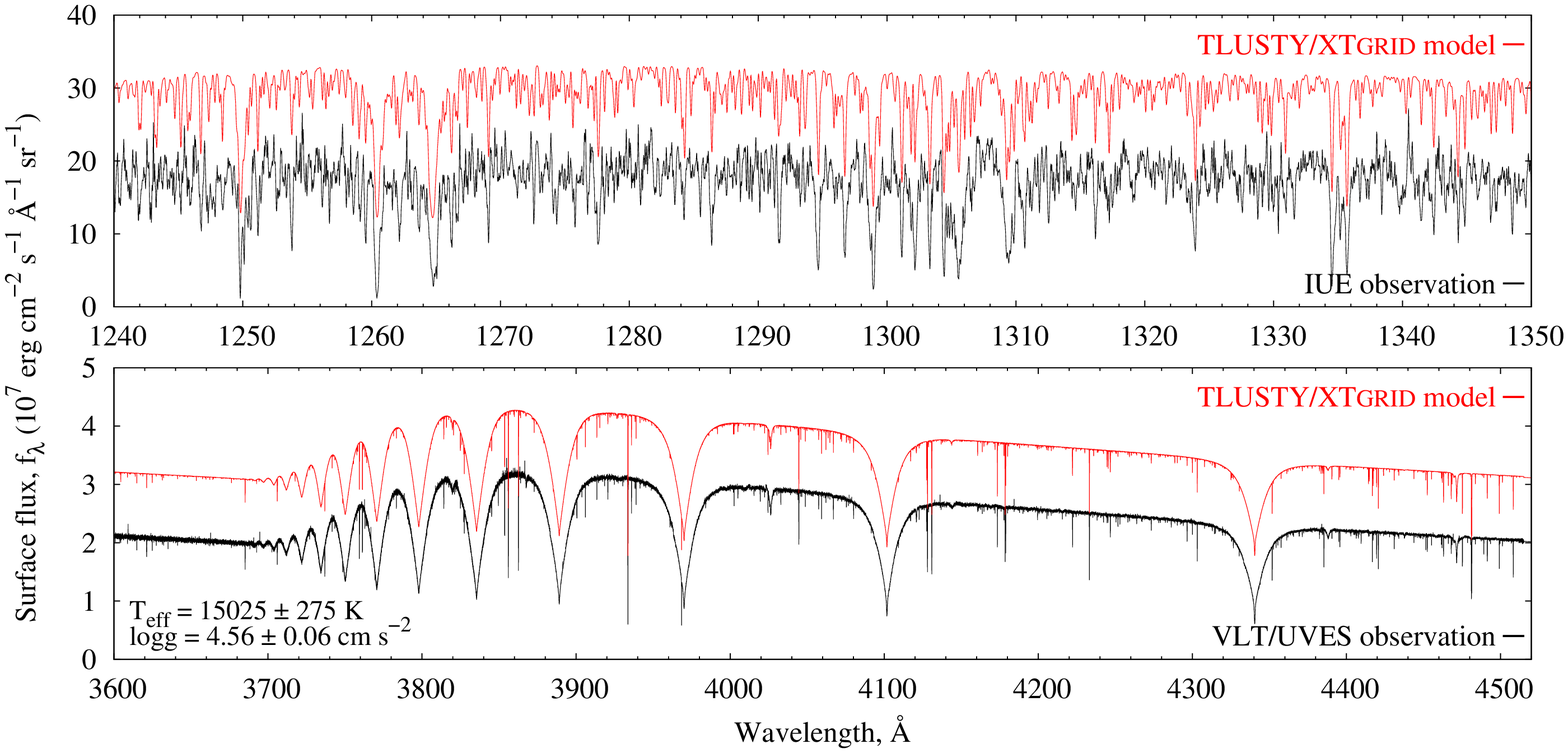}{fig3}{UV+O spectra of Feige 86 compared to a {\sc Tlusty/XTgrid} model. 
}
A precise parameter determination requires the knowledge of the radiation field as well as the composition and structure of the stellar atmosphere. Because the spectral energy distributions of the majority of stars peak in the UVOIR spectral range, a combination of data from different instruments are necessary to derive accurate surface parameters. Fig. \ref{fig3} shows such a fit for the blue horizontal branch (BHB) star Feige\,86 \citep{nemeth17a}. The spectrum of Feige\,86 shows clear signatures of atomic diffusion.

\subsection{Peculiar chemical compositions}
\articlefigure[width=\textwidth]{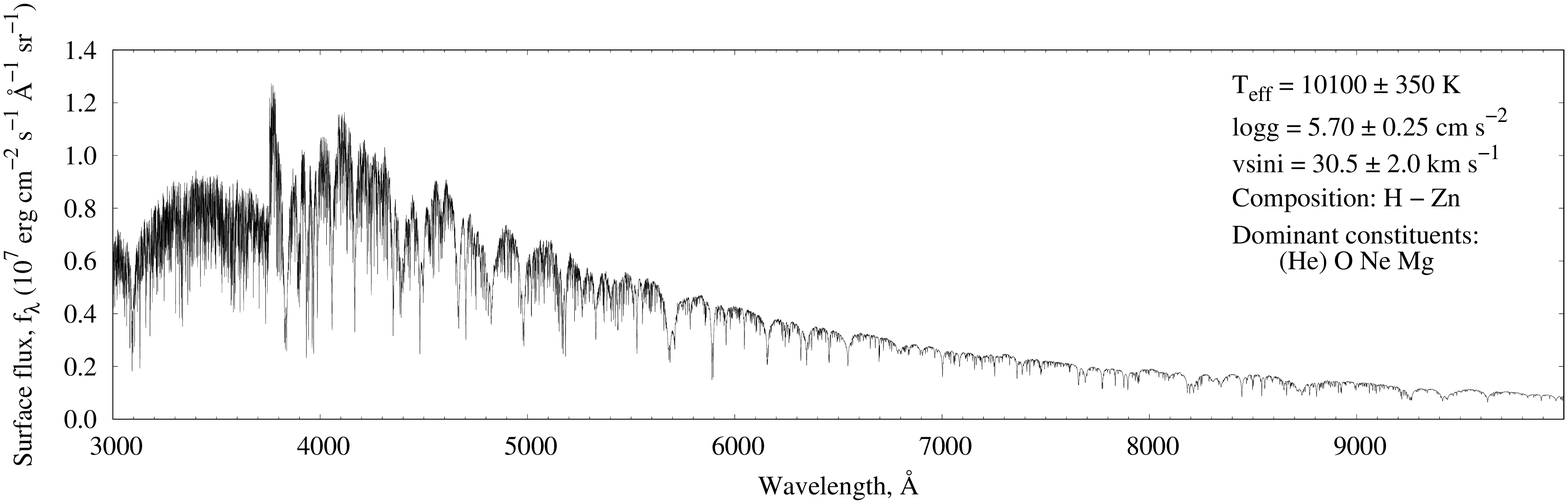}{fig4}{{\sc Tlusty/Synspec} model for LP\,40-365. This model for the atmosphere of the hypervelocity supernova remnant is dominated by metals.}
The complexity of flux redistribution due to non-LTE radiative transfer in metal rich atmospheres needs lengthy calculations with all major opacity sources present. The partially burnt supernova remnant LP\,40-365 \citep{vennes17} is an extremely metal-rich low-mass hypervelocity cool white dwarf. The best fitting synthetic spectrum in Fig. \ref{fig4} is a good example for such complex applications of {\sc XTgrid}.

\section{Summary}
The conference appointed future avenues for the field: New atomic data need to be implemented in model atoms and the performance of the analyses needs further improvements. The rate of incoming spectra with the future multi-object survey spectrographs will allow only seconds to fit the observations, which is currently extremely demanding.

The development and public release of {\sc Tlusty} marked important contributions to quantitative stellar astrophysics. {\sc Tlusty} helped the field become a cornerstone of modern astronomy. 
While future improvements will eventually polish a gem of that stone, we must also focus on its applications. 
To promote the use of {\sc Tlusty} we developed a web-service for {\sc XTgrid} and made it available through the sandbox services of Astroserver.org\footnote{\url{www.xtgrid.astroserver.org/sandbox/}} \citep{nemeth17b}. 
Beyond a continuous maintenance of the program our goal is to integrate all capabilities of {\sc Tlusty} into {\sc XTgrid}. 
Future developments will include a revision of the atomic data input as well as new analysis techniques for stratified atmospheres, irradiation in close binaries (extending on \citealt{vuckovic16}) and accretion disks. Such new methods will be added to the web-service as they become available.
Visit \url{www.Astroserver.org}
for news, more examples, and further references on how {\sc Tlusty} and {\sc XTgrid} may contribute to your research.

\acknowledgements My sincere gratitude goes to Ivan Hubeny, Thierry Lanz and all their collaborators for developing and distributing {\sc Tlusty/Synspec}. We thank Ad\'ela Kawka and St\'ephane Vennes for useful discussions, and acknowledge a generous support from the Grant Agency of the Czech Republic (GA\v{C}R 18-20083S).


\end{document}